\documentclass[12pt,preprint]{aastex}
\newcommand{\eq}{equation}
\newcommand{\eqn}{eqnarray}
\newcommand{\ds}{\displaystyle}

\makeatletter
\def\lsim{\mathrel{\mathpalette\gl@align<}}
\def\gsim{\mathrel{\mathpalette\gl@align>}}
\def\gl@align#1#2{\lower.6ex\vbox{\baselineskip\z@skip\lineskip\z@
    \ialign{$\m@th#1\hfil##\hfil$\crcr#2\crcr\sim\crcr}}}
\makeatother
\begin{document}

\title{Global Anisotropy Versus Small-Scale Fluctuations in
Neutrino Flux in Core-Collapse Supernova Explosions}

\author{Hideki Madokoro}
\email{madokoro@postman.riken.go.jp}

\author{Tetsuya Shimizu}
\email{tss@postman.riken.go.jp}

\and

\author{Yuko Motizuki}
\email{motizuki@riken.go.jp}

\affil{RIKEN, Hirosawa 2-1, Wako 351-0198, Japan}

\begin{abstract}
Effects of small-scale fluctuations in the neutrino radiation on core-collapse
supernova explosions are examined.
Through a parameter study with a fixed radiation field of neutrinos,
we find substantial differences between
the results of globally anisotropic neutrino radiation and those with
fluctuations.  As the number of modes of fluctuations increases, the shock
positions, entropy distributions, and explosion energies approach those of
spherical explosion.  We conclude that global anisotropy of the neutrino
radiation is the most effective mechanism of increasing the explosion energy
when the total neutrino luminosity is given.  This supports the previous
statement on the explosion mechanism by Shimizu and coworkers.

\end{abstract}


\clearpage

\section{INTRODUCTION}

For many years since the first work of \citet{CoWh66}, numerical simulations
of core-collapse supernova explosions have been exciting topics.  Until the
beginning of the 1990s, almost all the simulations included the assumption of
spherical symmetry \citep{Wi85}.  Such one-dimensional simulations, however,
were unable
to explain the observed explosion energy and often failed to produce explosions
(see, e.g., \citet{Lie01}).  Inclusion of convective motion via the mixing
length theory cures this problem to some extent \citep{WiMa93,BuMeDi95}.
However, spherical simulations based only on the Rayleigh-Taylor
instability require considerably large initial fluctuations in density to 
explain the large-scale matter mixing.  In addition, aspherical explosion is
also supported by the observation of SN1987A, where asymmetric ejecta are
clearly observed \citep{Wa02}.  These lead us to multidimensional simulations
of supernova explosions.

At this time, the two- and three-dimensional simulations have been performed
by several groups
\citep{MiWiMa93,HeBeHiFrCo94,BuHaFr95,JaMue96,Me98,FrHe00,ShEbSaYa01,FrWa02,
KiPlJaMu03}.  In many multidimensional simulations, special attention is paid
to the role of convection either near the surface of a nascent neutron star or
in neutrino-heated regions above the neutrinosphere.  It has been shown that
large-scale mixing, caused by convection and convective overturn around the
neutrino-heated region, increases the explosion energy and can trigger a
successful explosion \citep{HeBeHiFrCo94,KeJaMue96,JaMue96}.

Because supernova progenitors such as OB stars are generally observed
to be fast rotators ($\sim 200$ km s$^{-1}$ at the surface, $P\sim 1$ day;
see, e.g., Tassoul 1978; Fukuda 1982), the
resulting proto$-$neutron star can have a large amount of angular momentum
after the gravitational collapse.  Centrifugal force then deforms the rotating
core into an oblate form.  This will cause asymmetric neutrino radiation, in
which the flux along the pole is enhanced over that on the equatorial plane.
\citet{JaMoe89} first discussed the possibility of aspherical neutrino
emission from a rapidly rotating inner core.  They intended to evaluate the
total neutrino energy outputs using the neutrino data of SN~1987A detected
with the Kamiokande and Irvine-Michigan-Brookhaven experiments.  It was argued
in their paper that a neutrino flux
along the pole might be up to a factor of 3 greater than that on the
equatorial plane.

Inspired by this work, Shimizu and coworkers \citep{ShYaSa94,ShEbSaYa01} proposed
that the anisotropic neutrino radiation should play a critical role in the
explosion mechanism itself, and carefully investigated the effects of
anisotropic neutrino radiation on the explosion energy.  They found that only
a few percent enhancement in the neutrino emission along the pole is
sufficient to increase the explosion energy by a large factor, and that this
effect saturates around a certain degree of anisotropy.  It should be
noted here that the assumed rotational velocity of the inner core is very
different between \citet{JaMoe89} and Shimizu et al. (1994, 2001).
\citet{ShEbSaYa01} concluded that the increase in the explosion energy due
to anisotropic neutrino radiation occurs because cooling due to neutrino
reemission is suppressed in anisotropic models.  This is due to earlier shock
revival and hence more efficient decrease of the matter temperature than those
in spherical models.  On the other hand, the neutrino heating itself is almost
unchanged by the effect of anisotropic neutrino radiation.  The neutrino
heating dominates the cooling as a result, which increases the
explosion energy, and leads to a successful explosion.

In \citet{ShEbSaYa01}, the geometric effects of neutrino radiation have
been rigorously treated outside the neutrinosphere for the first time,
although its flux on the neutrinosphere was assumed.
Only a global form of anisotropy was
assumed there; the maximum peak in the neutrino flux distribution was
located at the pole, and the minimum at the equatorial plane.  However,
Burrows et al. (1995) have suggested that the neutrino flux can fluctuate with
angle and time.  Such fluctuations are due to gravitational oscillation on the
surface of the proto$-$neutron star and have a completely different origin from
that of globally anisotropic neutrino radiation.  Thus, it is interesting to
investigate how the small-scale fluctuations affect the explosion mechanism
and compare the results with those of the global anisotropy.

In this paper, we therefore introduce fluctuations in the neutrino flux in our
numerical code by modifying the angular distribution of the neutrino flux.
We aim to study the effects of these small-scale fluctuations on the shock
position, the explosion energy, and the asymmetric explosion.  Our numerical
simulation is described in \S~2, and the results are presented and discussed
in \S~3.  Our conclusion is given in \S~4.

\section{NUMERICAL SIMULATION}
Our simulation is performed by solving two-dimensional hydrodynamic equations
in spherical coordinates.  A generalized Roe's method is employed to solve the
hydrodynamic equations with general equations of state (EOSs). The details of
our numerical technique, together with the EOS and the initial condition used,
are described in the previous article \citep{ShEbSaYa01}.  In our study,
we have improved the numerical code of \citet{ShEbSaYa01};  the cells in
the $\theta$-direction were shifted by half of the cell size \citep{Sh95} in
order to avoid a numerical error near the pole, although the error was not
serious for the investigation of the explosion energy.  The computational
region is divided into 500 ($r$-direction) $\times$ 62 ($\theta$-direction)
numerical cells.

In the present paper, the local neutrino flux is assumed to be
\begin{\eq}
  l_{\nu}(r,\theta) = \frac{7}{16}\sigma T_{\nu}^{4} c_{1}
  \left(1+c_{2}\cos^{2}(n_{\theta}\theta)\right)\frac{1}{r^{2}},
  \label{eqn:nuflux}
\end{\eq}

\noindent
where $\sigma$ is the Boltzmann constant, and $T_{\nu}$ is the temperature on
the neutrinosphere.  In equation (\ref{eqn:nuflux}), the parameter $c_{2}$
represents the magnitude of anisotropy, and $n_{\theta}$ the number of waves in
the $\theta$-direction.  The case of $n_{\theta}=1$ corresponds to the global
anisotropy, namely, no fluctuation.  We see in equation (\ref{eqn:nuflux})
that the neutrino fluxes in the $x$ (equatorial) and $z$ (polar) directions
become $l_{x} \equiv l_{\nu}(r,\theta=90^{\circ}) \propto c_{1}$ and $l_{z}
\equiv l_{\nu}(r,\theta=0^{\circ}) \propto c_{1}(1+c_{2}),$ respectively.
The degree of anisotropy $l_{z}/l_{x}$ is then represented as
\begin{\eq}
  \frac{l_{z}}{l_{x}} = 1 + c_{2}.
  \label{eqn:anisotropy}
\end{\eq}

\noindent
Note that equation (\ref{eqn:anisotropy}) is different from that defined
by \citet{ShEbSaYa01}, $(l_{z}/l_{x})_{\rm Shimizu}$; for $n_{\theta}=1$ and
sufficiently small $c_{2}$, we can relate them as $c_{2} \sim
\left[(l_{z}/l_{x})_{\rm Shimizu}^{2}-1\right]/2$.

The value of $c_{1}$ is calculated from given $c_{2}$ and $n_{\theta}$ so as
to adjust the total neutrino flux to that in the spherical model.  The total
neutrino luminosity is obtained by integrating equation (\ref{eqn:nuflux})
over the whole solid angle,
\begin{\eq}
  L_{\nu} = \int r^{2}l_{\nu}(r,\theta) d\Omega
          = \frac{7}{16}\sigma T_{\nu}^{4}\;4\pi c_{1}
            \left(1+c_{2}\;\frac{2n_{\theta}^{2}-1}{4n_{\theta}^{2}-1}\right),
  \label{eqn:lnutotal}
\end{\eq}

\noindent
which is equated to that of spherical explosion with the same $T_{\nu}$,
\begin{\eq}
  L_{\nu}^{\rm sp} = \frac{7}{16}\sigma T_{\nu}^{4}\;4\pi
  R_{\rm NS}^{2}.
  \label{eqn:lnutotalsph}
\end{\eq}

\noindent
In the above, $R_{\rm NS}$ is the radius of a proto$-$neutron star and fixed to
be 50 km.  By comparing equations (\ref{eqn:lnutotal}) with
(\ref{eqn:lnutotalsph}), we obtain
\begin{\eq}
  c_{1}=\frac{R_{\rm NS}^{2}}{1+c_{2}\;
        (2n_{\theta}^{2}-1)/(4n_{\theta}^{2}-1)}.
\end{\eq}

It should be noted here that the magnitude of fluctuations in the neutrino
flux distribution for an observer far from the neutrinosphere (represented by
$c_{2}$) and that on
the neutrino-emitting surface (here we denote it as $a$) are different:
the local neutrino flux is seen as equation (\ref{eqn:nuflux}) when we observe
fluctuations on the surface of neutrino emission far from the neutrinosphere.
It is preferable that we compare the results for the same value of $a$, since
$a$ is more directly related to explosion dynamics.  The value of $c_{2}$
should, therefore, be calculated from a given $a$, depending on $n_{\theta}$.
Although it is difficult to calculate the exact relationship between $c_{2}$
and $a$, we can estimate it as follows.  First, we assume that the strength
of the neutrino flux on the neutrinosphere is represented by a profile of
step functions:
\begin{\eq}
  \begin{array}{lll}
    \mbox{for }n_{\theta}=1 \\
    & f(\theta) & \propto \left\{
      \begin{array}{l}
        1+a \;\;\left(1 \geq \cos\theta > 1/2\right) \\
        1 \;\;\left(1/2 > \cos\theta \geq 0\right),
      \end{array} \right. \\
    \mbox{for }n_{\theta}=3 \\
    & f(\theta) & \propto \left\{
      \begin{array}{l}
        1+a \;\;\left(1 \geq \cos\theta > 3/4,\;1/2 >
        \cos\theta > 1/4\right) \\
        1 \;\;\left(3/4 > \cos\theta > 1/2,\;
      1/4 > \cos\theta \geq 0)\right),
      \end{array} \right. \\
    \mbox{for }n_{\theta}=5 \\
    & f(\theta) & \propto \left\{
      \begin{array}{l}
        1+a \;\;\left(1 \geq \cos\theta > 5/6,\;2/3 >
        \cos\theta > 1/2,\;1/3 > \cos\theta > 1/6
        \right) \\
        1 \;\;\left(5/6 > \cos\theta > 2/3,\;
        1/2 > \cos\theta > 1/3,\;1/6 > \cos\theta \geq
        0\right).
      \end{array} \right.
  \end{array}
  \label{eqn:fluxonns}
\end{\eq}
\noindent
We continue similarly for larger values of $n_{\theta}$.  The parameter $a$
in equation (\ref{eqn:fluxonns}) represents the magnitude of fluctuations
in the neutrino flux on the neutrino emitting surface; the bright regions
($f\sim 1+a$) correspond to those where rising convective motion occurs,
while the dark regions ($f\sim 1$) correspond to those of sinking convection.
We have assumed that the areas of bright and dark regions are the same.
Note that the neutrino flux function is always bright ($f\sim 1+a$) when
$\theta=0$ (along the pole) and dark ($f\sim 1$) in the case of $\theta=\pi/2$
(on the equatorial plane), and that the fluctuations are essentially added to
the global anisotropic model ($n_{\theta}=1$).

The neutrino flux observed far from the neutrinosphere is obtained by
averaging all contributions from the flux on the surface of the neutrinosphere,
$l(\Theta)=\int d\phi\int d\theta f(\theta) \sin\theta \cos(\theta-\Theta)$.
Here $\Theta$ is the inclination angle between the line of sight of an
observer and the polar axis of the proto$-$neutron star.  This means that fully
geometric effects from an anisotropically radiating surface are included in
neutrino radiation field.  For example, there is a contribution from fluxes
along the pole axis to those on the equatorial plane, which will tend to
reduce an efficiency of anisotropy.  The ratio of the local neutrino flux
along the polar axis ($l_{z}$) to that on the equatorial plane ($l_{x}$) for
an observer far from the neutrinosphere is described as
\begin{\eqn}
  \frac{l_{z}}{l_{x}} & = &
      \frac{\ds \int_{0}^{2\pi}d\phi
      \int_{0}^{\pi/2}f(\theta) \sin\theta \cos\theta d\theta}
      {\ds \int_{0}^{\pi}d\phi \int_{0}^{\pi}
      f(\theta) \sin\theta \cos\left(\theta-\pi/2\right) d\theta}
\nonumber \\
  & = & \left\{
  \begin{array}{l}
    \frac{\ds 1+0.750 a}{\ds 1+0.391 a}\;\;\mbox{for $n_{\theta}=1$},
      \vspace{2ex} \\
    \frac{\ds 1+0.625 a}{\ds 1+0.438 a}\;\;\mbox{for $n_{\theta}=3$},
      \vspace{2ex} \\
    \frac{\ds 1+0.583a}{\ds 1+0.457 a}\;\;\mbox{for $n_{\theta}=5$}, \cdots
  \end{array}\right.
  \label{eqn:lxlz2}
\end{\eqn}

\noindent
By comparing equation (\ref{eqn:lxlz2}) with equation (\ref{eqn:anisotropy}),
we finally obtain the value of $c_{2}$ for each value of $n_{\theta}$.

In the following, we examine two model series: $a=0.31$ (model series A) and
$a=0.71$ (model series B).  These values of $a$ are chosen in such a way that
the value of $l_{z}/l_{x}$ for the global model ($n_{\theta}=1$) becomes 1.10
and 1.20, respectively.  The values of $c_{2}$ for each fluctuation model
($n_{\theta}=3,5$) are accordingly calculated.  These are summarized in
Table~{\ref{tab:models}}.  The neutrino temperature on the neutrino-emitting
surface $T_{\nu}$ is assumed to be 4.65 and 4.70 MeV.  Note that the value
of 1+$a$, 1.71, for the model series B roughly corresponds to the variation
in the neutrino flux obtained by Burrows et al. (1995), which is at a factor
of 1.6.

\section{RESULTS AND DISCUSSION}

\begin{figure}
%\plottwo{c2=0105_nt=1_082ms.ps}{c2=0105_nt=1_244ms.ps}
\plottwo{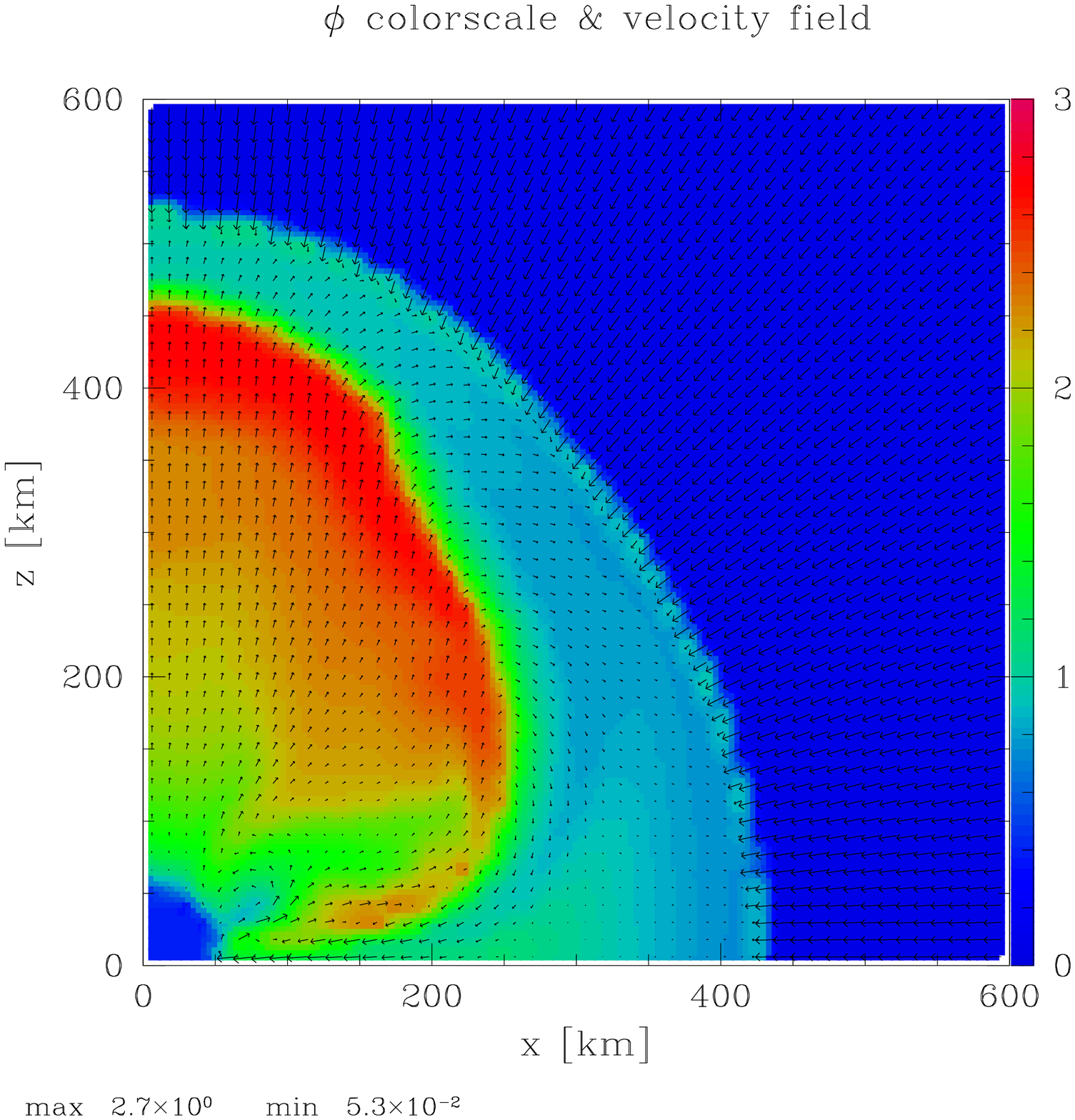}{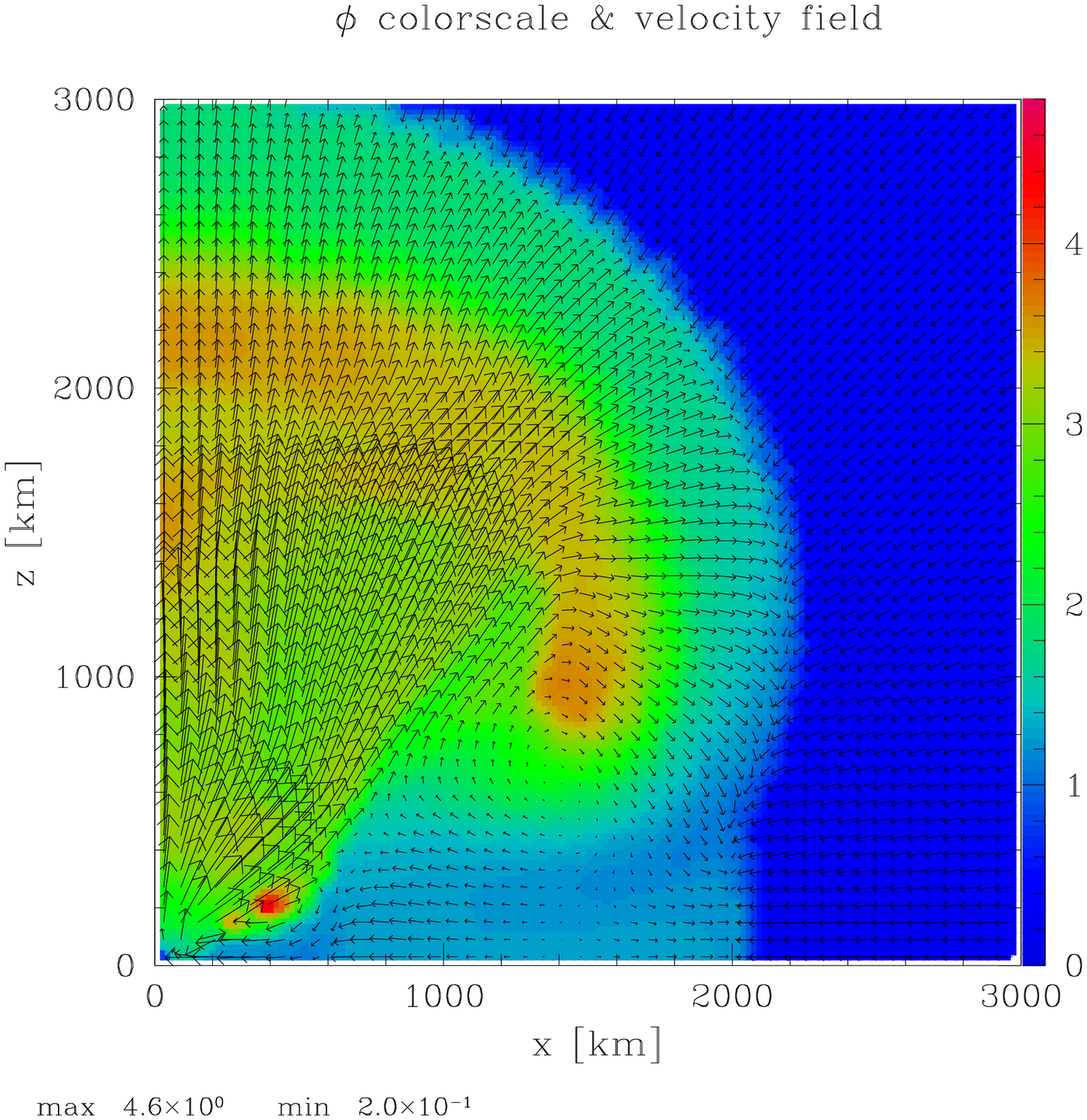}
\figcaption{Color-scale maps of the dimensionless entropy distribution and
           the velocity fields for the model of $n_{\theta}=1$ and
           $T_{\nu}=4.70$ MeV of series A (model A1-T470).
           Left: $t=82$ms after the shock stall, Right: $t=244$ms.
           \label{fig:entropy470_global_A}}
\end{figure}

\begin{figure}
%\plottwo{c2=0200_nt=1_082ms.ps}{c2=0200_nt=1_249ms.ps}
\plottwo{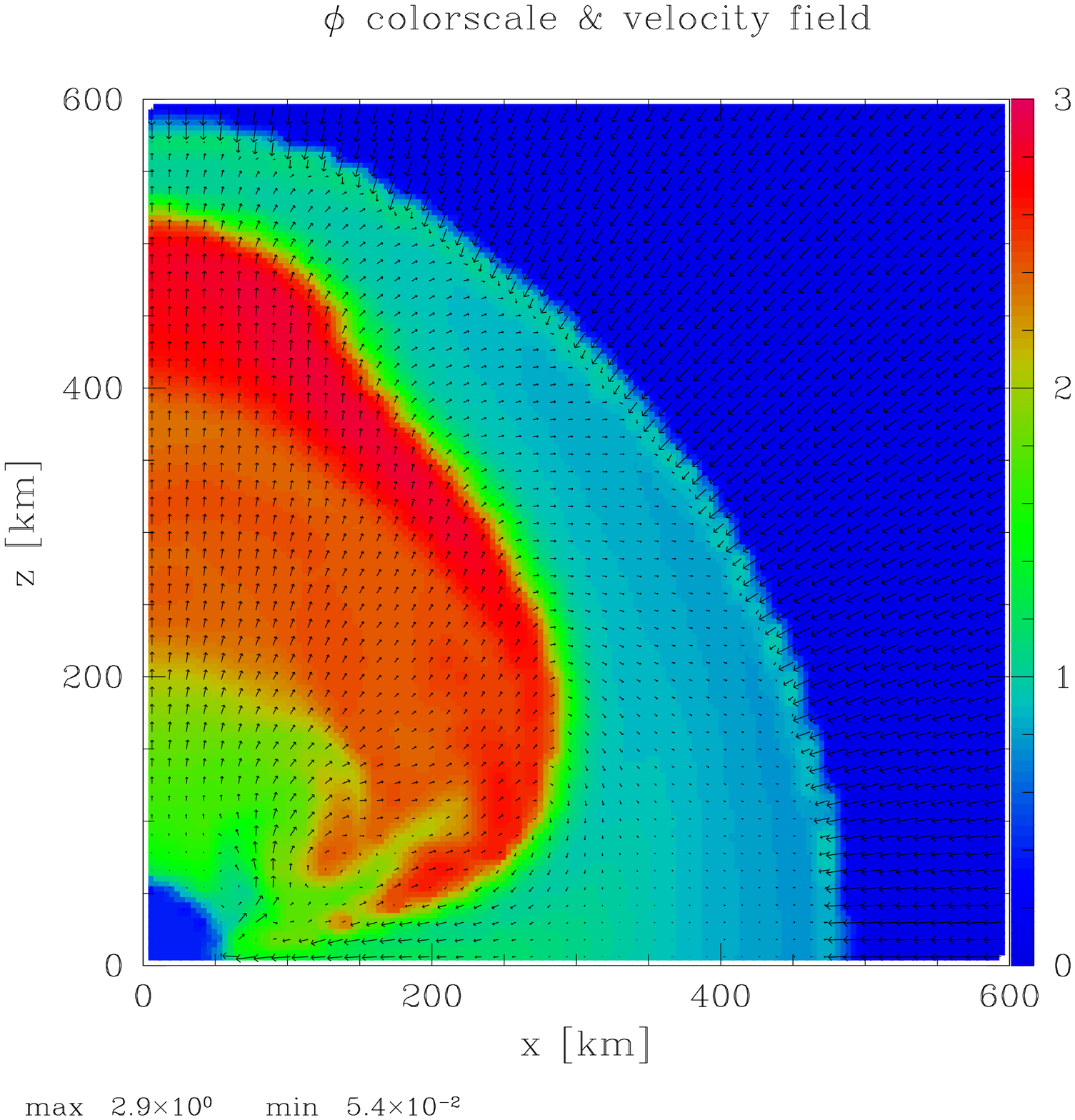}{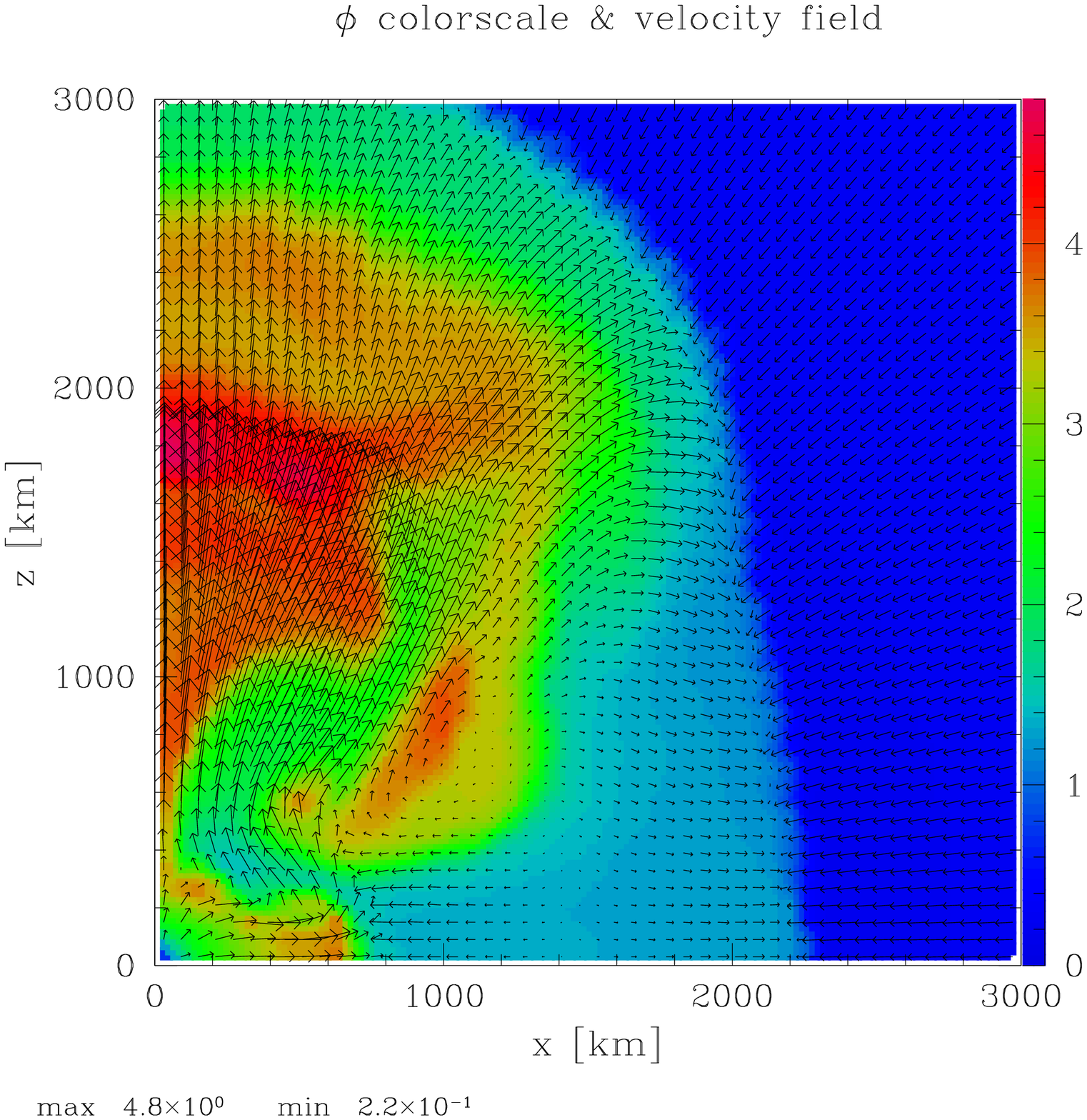}
\figcaption{Same as Fig.\ref{fig:entropy470_global_A}, except for series B
           (model B1-T470).  Left: $t=82$ms after the shock stall, Right:
           $t=249$ms.
           \label{fig:entropy470_global_B}}
\end{figure}

Figures~\ref{fig:entropy470_global_A} and ~\ref{fig:entropy470_global_B} show
the color-scale maps of the dimensionless entropy \citep{ShEbSaYa01}
distribution with the velocity fields for the model of global anisotropy
($n_{\theta}=1$; models A1-T470 and B1-T470).  At $t=82$ ms after the shock
stall, the shock front reaches $r\sim 430$ km on the equatorial plane and
$r\sim 530$ km at the pole for the model A1-T470.  The shock front is prolate,
since the neutrino heating along the pole is more intensive than that on
the equatorial plane, resulting in a jetlike explosion.  At a later stage
($t=244$ ms), the shock wave is around a few thousand kilometers with
large distortion.  The entropy distribution of the model B1-T470 has a similar
profile except that the shock front is more extended.

\begin{figure}
%\plottwo{c2=0051_nt=3_254ms.ps}{c2=0035_nt=5_250ms.ps}
\plottwo{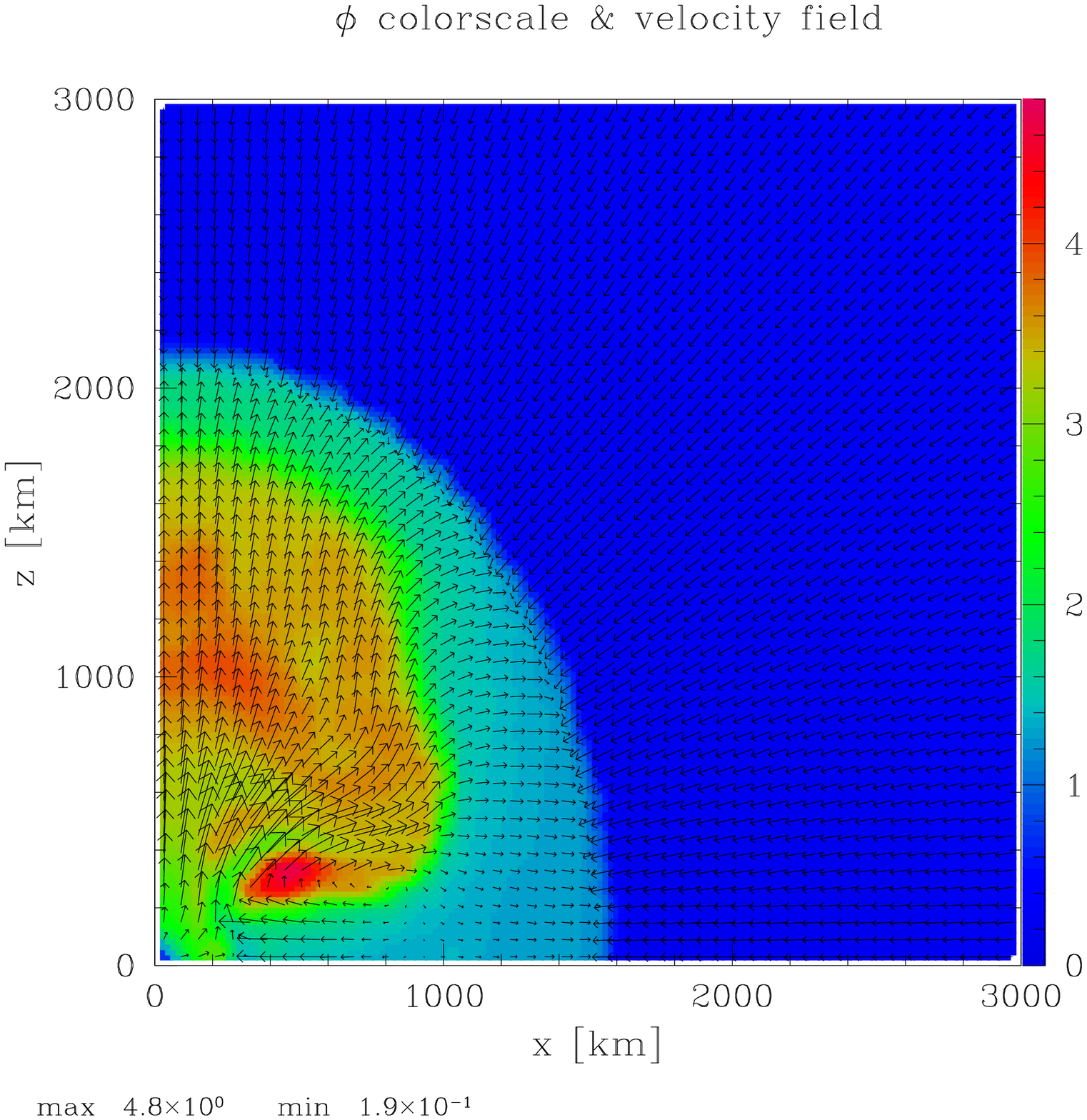}{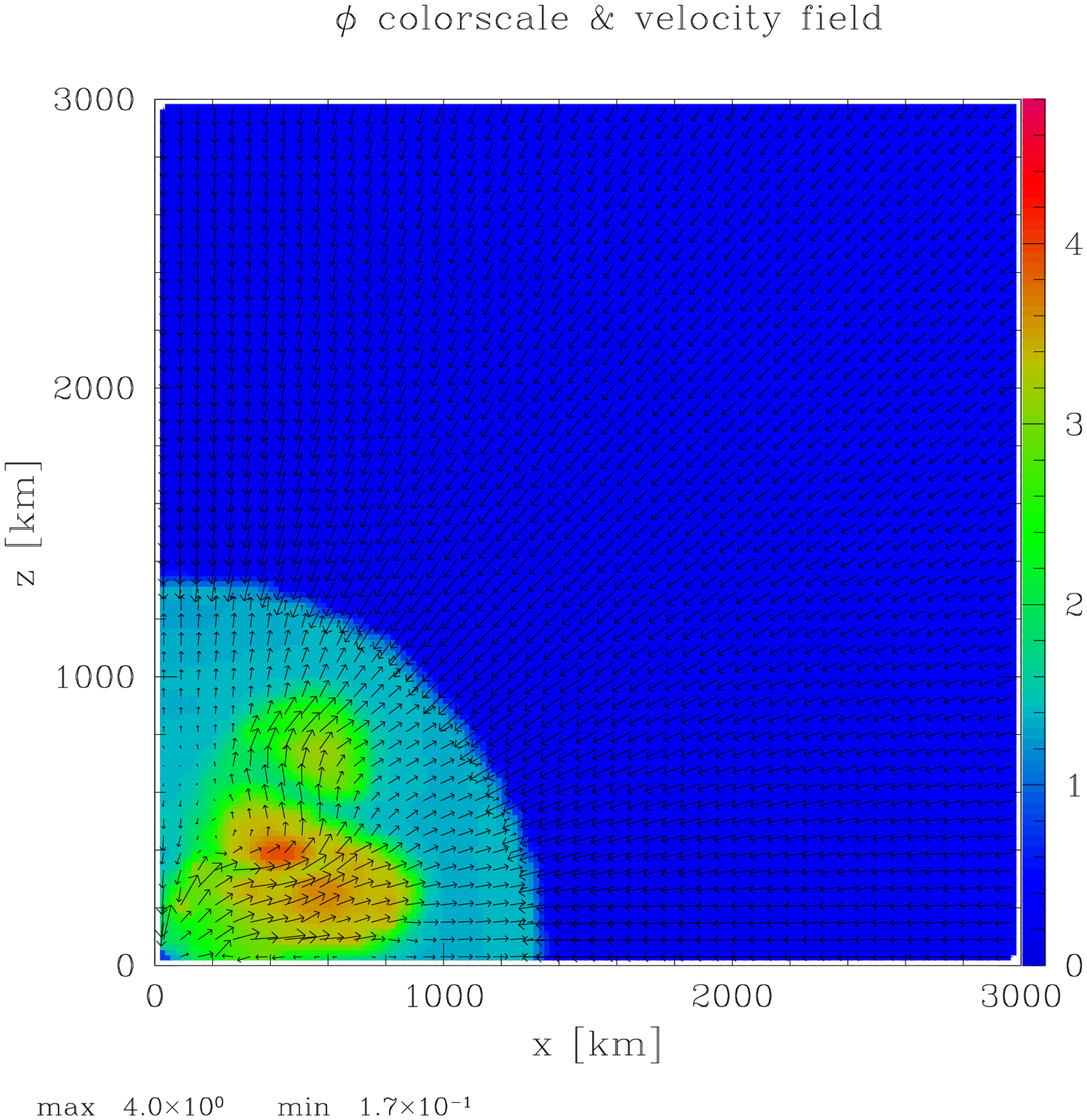}
\figcaption{Same as Fig.\ref{fig:entropy470_global_A}, except for the case of
           fluctuated neutrino flux.  Left: $n_{\theta}=3$ (model A3-T470)
           at $t=254$ms after the shock stall, Right: $n_{\theta}=5$
           (model A5-T470) at $t=250$ms.
           \label{fig:entropy470_fluct_A}}
\end{figure}

\begin{figure}
%\plottwo{c2=0105_nt=3_257ms.ps}{c2=0068_nt=5_247ms.ps}
\plottwo{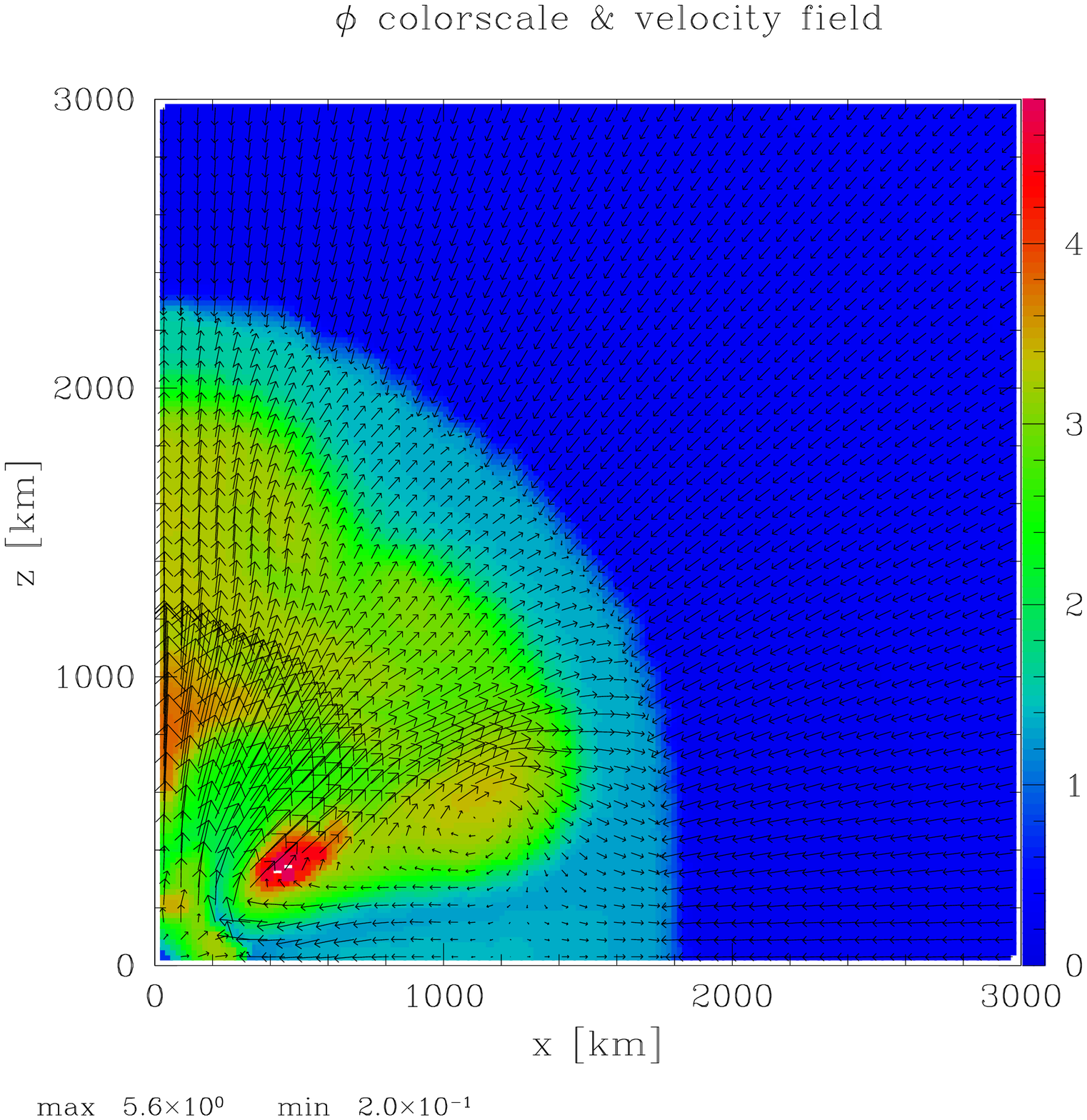}{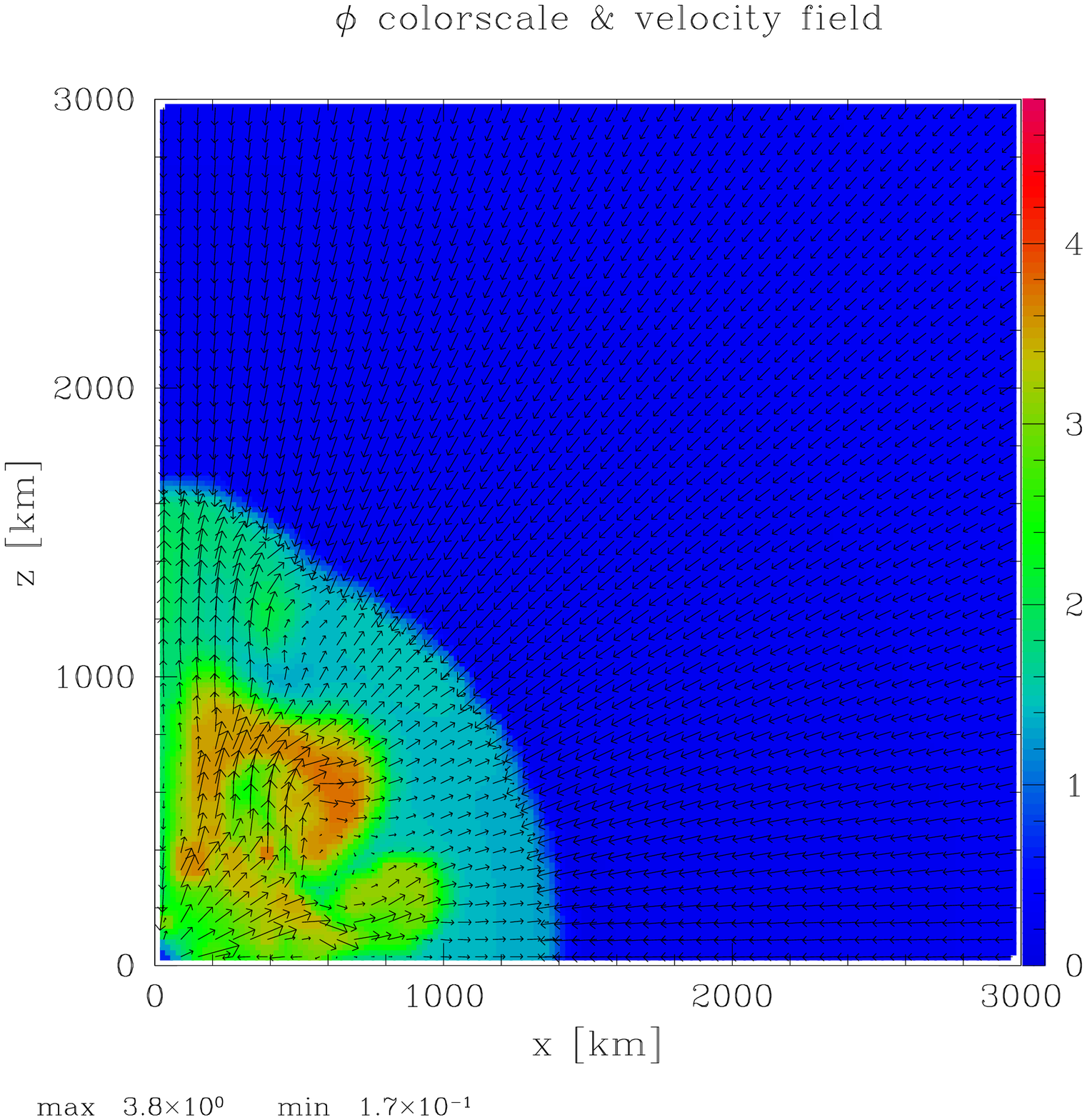}
\figcaption{Same as Fig.\ref{fig:entropy470_fluct_A}, except for series B.
            Left: $n_{\theta}=3$ (model B3-T470) at $t=257$ms after
            the shock stall, Right: $n_{\theta}=5$ (model B5-T470) at
            $t=247$ms.
           \label{fig:entropy470_fluct_B}}
\end{figure}

In Figures~\ref{fig:entropy470_fluct_A} and ~\ref{fig:entropy470_fluct_B},
the results for models with fluctuations (models A3-T470, A5-T470, B3-T470,
and B5-T470) are depicted.  When $n_{\theta}=3$, we find that the degree of
asymmetry is smaller than that of the global anisotropy.  In particular,
the shock position for these models is less extended than
the globally anisotropic one.  This trend becomes more obvious for the model
of $c_{2}=+0.035$ and $n_{\theta}=5$ (model A5-T470), where the shock front is
almost spherical and its radius is only about 1300 km.  In the case of
the model B5-T470, the shock front is distorted because of a strong
hydrodynamic
flow along the pole, although this does not affect the explosion energy
(see discussion on the energy later in this section).

\begin{figure}
%\plottwo{energies_a=03.ps}{energies_a=07.ps}
\plottwo{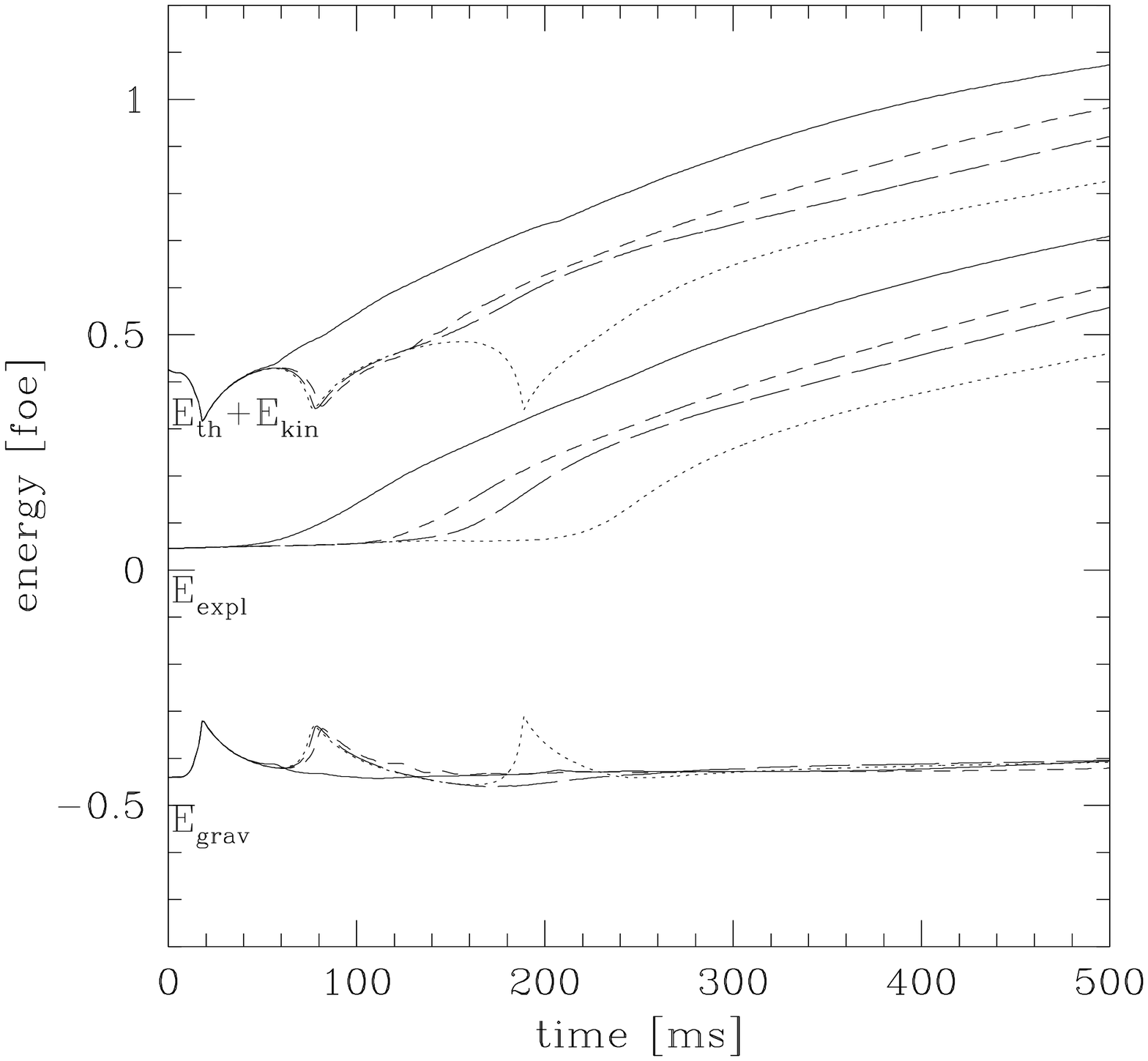}{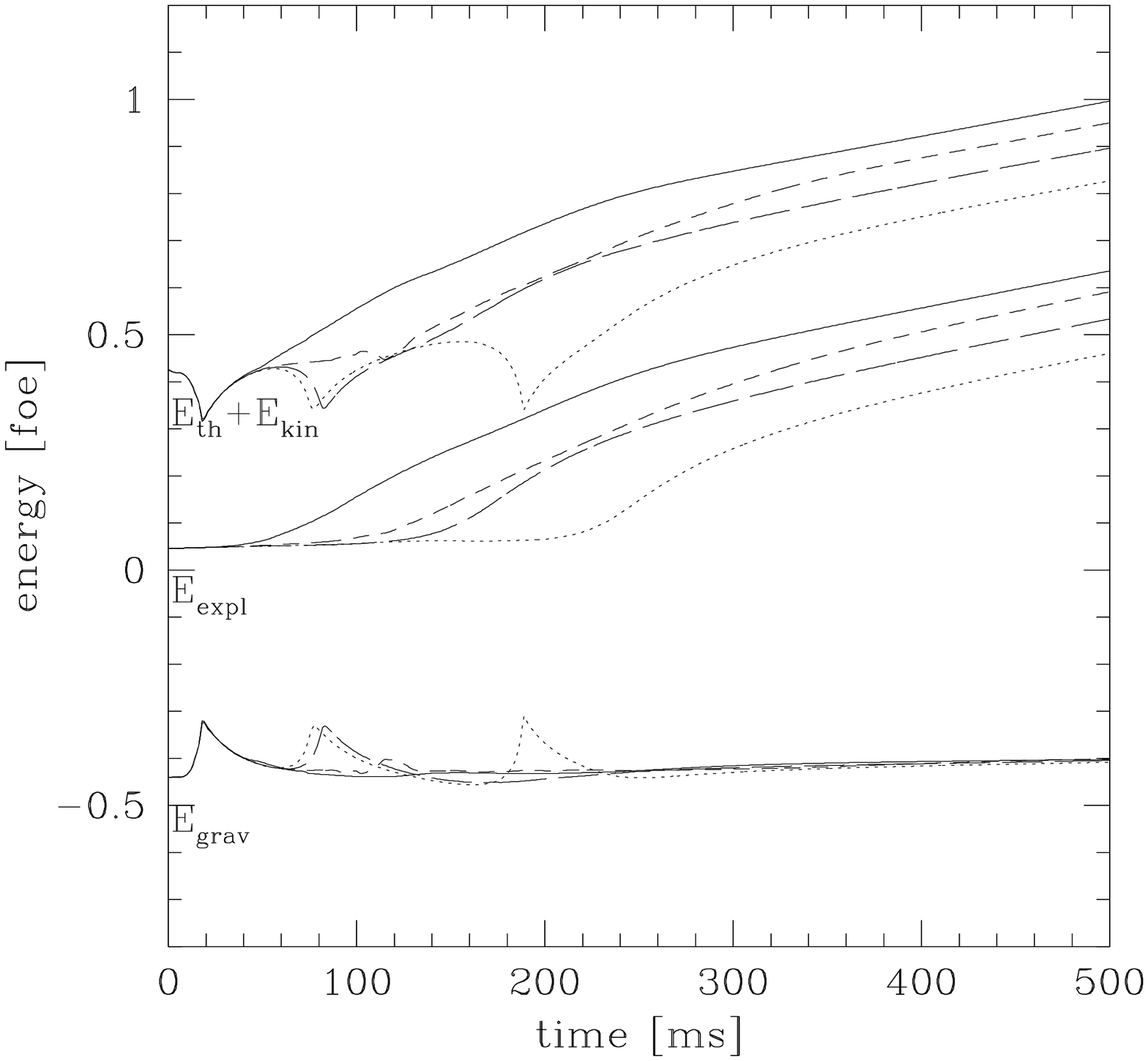}
\figcaption{Evolution of thermal and kinetic energy
           ($E_{\rm th}+E_{\rm kin}$), gravitational
           energy ($E_{\rm grav}$) and explosion energy ($E_{\rm expl}$)
           for the models of $T_{\nu}=$4.70 MeV.
           Solid line corresponds to the case of $n_{\theta}=1$, short-dashed
           line $n_{\theta}=3$, long-dashed line $n_{\theta}=5$, and
           dotted line $c_{2}=0$ (spherical).  Left: model series A, Right:
           model series B.
           \label{fig:energy470}}
\end{figure}

The profile of the explosion energy is found to be closer to that of
spherical explosion as the mode number of fluctuations increases.
Figure~\ref{fig:energy470} shows the evolution of the explosion energy, as
well as the thermal, kinetic, and gravitational energies for the models of
$T_{\nu}=4.70$ MeV.  The difference between the globally anisotropic model and
the models with fluctuations is prominent: the energy gain for the case of
$n_{\theta}=1$ is the highest among others at all stages of the explosion.
It is also seen that the explosion energy decreases as the mode number of
fluctuations in the neutrino flux increases and finally approaches that of
spherical explosion.  For series B, we have obtained similar results.
We observe that asymmetry in explosive motion is more enhanced than that for
each model of series A.

In Figure~\ref{fig:energy470}, we compare the results of the
explosion energy for the two model series.  No significant difference is
found between the two, although the result of the model A1-T470 becomes larger
than that of the model B1-T470 at the later stages of the explosion.  It has
been
shown \citep{ShEbSaYa01} that the explosion energy increases as the degree of
anisotropy becomes larger for not so large a degree of anisotropy, and finally
saturates at $(l_{z}/l_{x})_{\rm Shimizu}\sim 1.2$.  Our new result shows that
the final explosion energy of the more anisotropic model B1-T470 is smaller
than that of the model A1-T470 (see Fig. \ref{fig:energy470}).  The
difference may be attributed to the fact that the assumed forms of the local
neutrino flux are different (compare eq. [\ref{eqn:nuflux}] here with eq. [5]
in Shimizu et al. 2001).  In the present paper, we have assumed the form of
the neutrino fluxes that has more sharply concentrated flux on
the pole.  Therefore, the neutrino heating and rising convection are focused
on the pole, and those in the equatorial direction are extremely reduced.
The shock wave of the model B1-T470 appears to be too weak on the equatorial
plane at $t\gsim 300$ms, which causes a energy loss \citep{ShEbSaYa01}.
Such features are clearly seen in the entropy distribution of the model
B1-T470.

\begin{figure}
%\plottwo{energies_a=03_T465.ps}{energies_a=07_T465.ps}
\plottwo{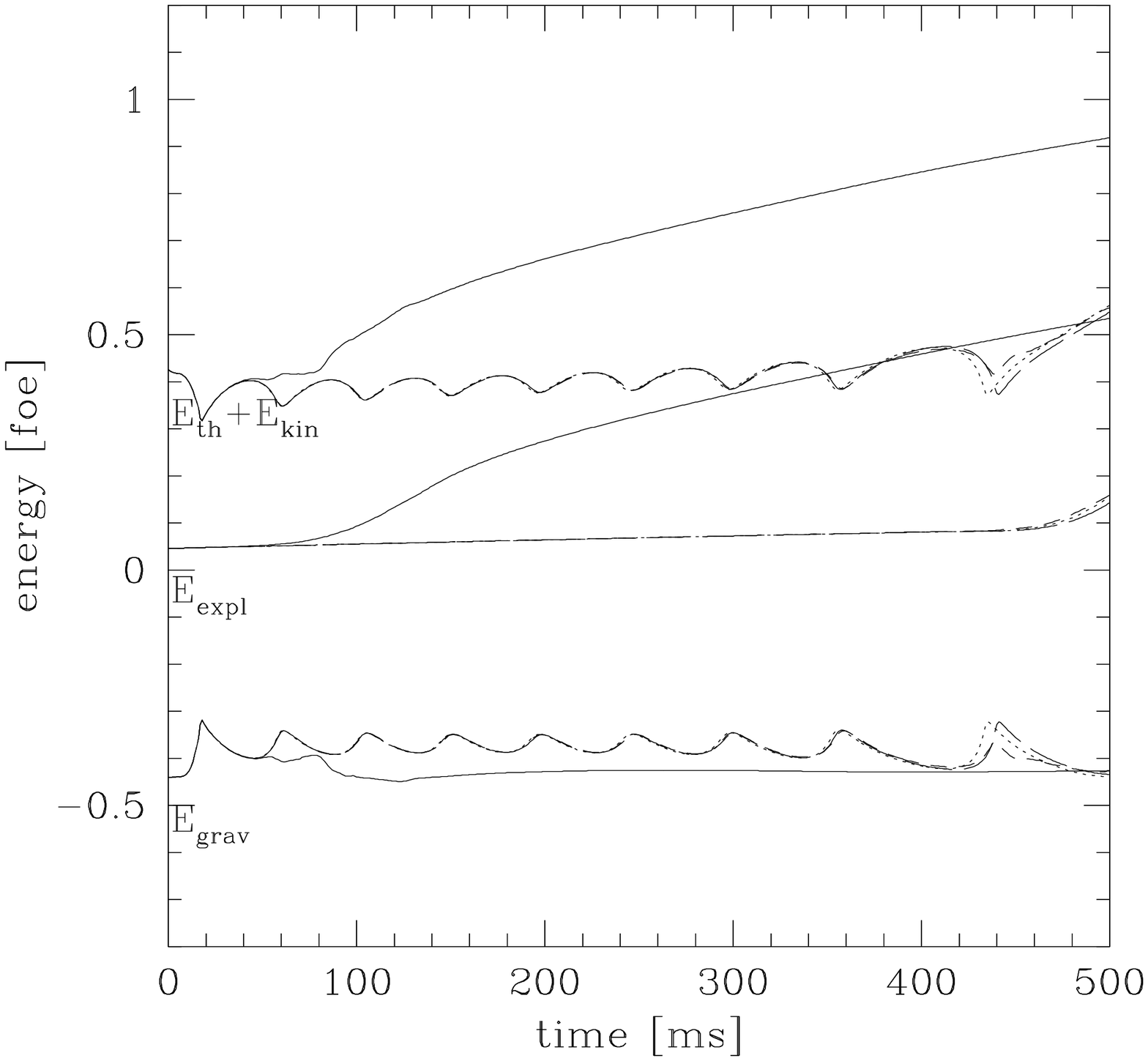}{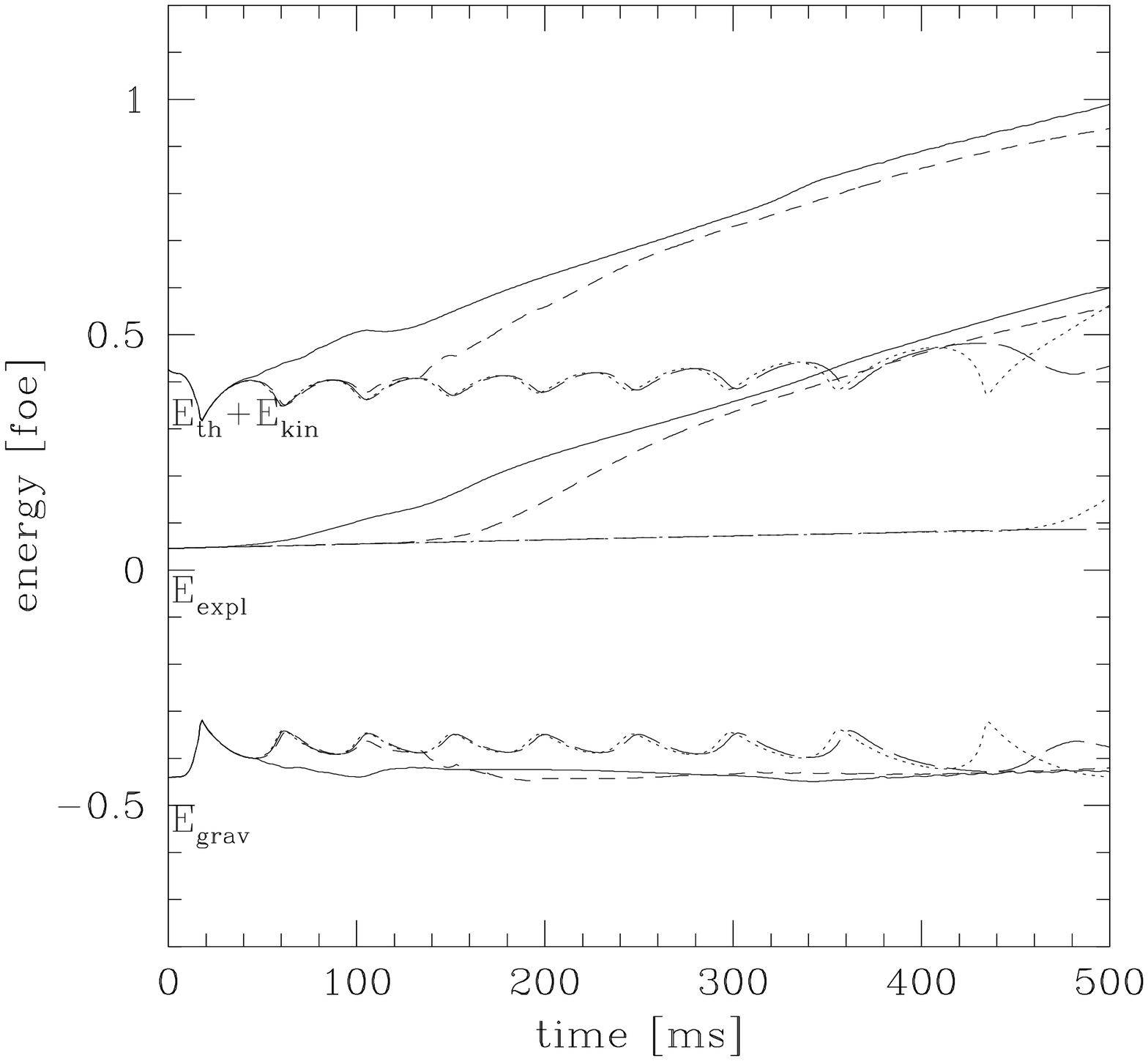}
\figcaption{Same as Fig.\ref{fig:energy470}, except for
            $T_{\nu}=4.65$ MeV.  Left: model series A, Right: model series B.
           \label{fig:energy465}}
\end{figure}

The effectiveness of global anisotropy becomes more pronounced as $T_{\nu}$ is
decreased.  Figure~\ref{fig:energy465} shows the same energy evolution as
Figure~\ref{fig:energy470}, except for $T_{\nu}=4.65$ MeV.  The difference
between the model with global anisotropy and those with fluctuations is
extremely remarkable.  The globally anisotropic models succeed, while all
the other fluctuated and spherical models fail to explode except for
the model B3-T465.
Note that difference in the neutrino temperature is only 1\%, which
indicates a sensitivity of the supernova problem on the neutrino
luminosity and energy.  Note also that an increase of
the explosion energies of the models with fluctuations at $t\sim 500$ ms is
physically meaningless, because we have not taken into account a decay of the
neutrino luminosity when $t\gsim 500$ ms (e.g., Wilson \& Mayle 1988; see also
Shimizu et al. 2001).

We found that there are remarkable differences in the explosion energy depending on
the mode of the fluctuations and that larger number of modes in the fluctuations
makes the result closer to that of spherical explosion, irrespective of
the model series A or B.  Any small-scale fluctuations on the neutrinosphere
are greatly averaged out when the neutrino emission is observed far enough from
the neutrino-emitting surface.  Moreover, we found that a certain broad space
is needed to be heated by neutrinos to revive the stalled shock wave rigorously
and that the global anisotropy ($n_{\theta}=1$) is the most effective
to increase the explosion energy.  Burrows et al. (1995) suggested that
the neutrino flux can fluctuate not only with angle but with time.  Such time
fluctuations are expected to reduce further the efficiency of anisotropy,
which needs to be confirmed in the future.

\section{CONCLUSION}

We have investigated the effects of small-scale fluctuations in the neutrino
flux on the core-collapse supernova explosion.  In order to examine the effect
of the degree of anisotropy itself on the explosion, we have studied two
model series parametrically.  We specified the neutrino radiation field
taking its geometric effects into account in each model.
We found that the global anisotropy
($n_{\theta}=1$) and the local fluctuations ($n_{\theta}>1$) in the neutrino
flux have quite different effects on the explosion mechanism, that is, the
shock dynamics, the explosion energy, and the explosion asymmetry.  Since the
small-scale fluctuations are averaged out for radiative and hydrodynamic
reasons, the results including fluctuations become closer to that of spherical
explosion.  Consequently, the global anisotropy is the most effective
mechanism of increasing the explosion energy.
Note here that the explosion energy could differ substantially depending on
the neutrino temperature (the difference between 4.70 and 4.65 MeV
is only 1\%).
This indicates that the supernova problem is very sensitive to the neutrino
energy and luminosity.  However, the total
luminosity cannot be simply increased to explain the observed explosion energy
because such treatment leads to the problem of Ni overproduction, especially
in the case of essentially spherical models.  We therefore conclude that
globally anisotropic neutrino radiation is of great importance in actual
supernova explosions.  This supports the claim made by \citet{ShEbSaYa01}.

The global anisotropy can originate from rotation of a proto$-$neutron
star or a hot spot on the neutrino-emitting region, while the small-scale
fluctuations are considered to be resulted from gravitational oscillation or
uniform convection.  It will be very interesting if any evidence of anisotropic
neutrino radiation is observed at facilities like Super Kamiokande
\citep{Hi87,Su98} and SNO \citep{Po01}, together with detailed optical
observations \citep{Wa02}.

\acknowledgments

We are grateful to the anonymous referee for useful comments and suggestions
that improved this paper.

\clearpage

\begin{deluxetable}{cccccc}
%\tabletypesize{\scriptsize}
\tablecaption{Simulated Models}
\tablewidth{0pt}
\tablehead{
  \colhead{Model Series} & \colhead{$a$\tablenotemark{a}} & \colhead{Model}
  & \colhead{$n_{\theta}$\tablenotemark{b}}
  & \colhead{$c_{2}$\tablenotemark{c}} & \colhead{$T_{\nu}$\tablenotemark{c}
    (MeV)}
}
  \startdata
       &      & A1-T465 & 1 & 0.100 & 4.65 \\
       &      & A1-T470 & 1 & 0.100 & 4.70 \\
    A  & 0.31 & A3-T465 & 3 & 0.051 & 4.65 \\
       &      & A3-T470 & 3 & 0.051 & 4.70 \\
       &      & A5-T465 & 5 & 0.035 & 4.65 \\
       &      & A5-T470 & 5 & 0.035 & 4.70 \\ \tableline
       &      & B1-T465 & 1 & 0.200 & 4.65 \\
       &      & B1-T470 & 1 & 0.200 & 4.70 \\
    B  & 0.71 & B3-T465 & 3 & 0.101 & 4.65 \\
       &      & B3-T470 & 3 & 0.101 & 4.70 \\
       &      & B5-T465 & 5 & 0.068 & 4.65 \\
       &      & B5-T470 & 5 & 0.068 & 4.70 \\% \hline
  \enddata
  \tablenotetext{a}{Magnitude of fluctuations on the neutrino emitting surface.}
  \tablenotetext{b}{Mode number of fluctuations.}
  \tablenotetext{c}{Magnitude of fluctuations far enough from the neutrino
    emitting surface.}
  \tablenotetext{d}{Temperature on the neutrinosphere}
  \label{tab:models}
\end{deluxetable}

\clearpage

\clearpage

\clearpage

\end{document}